\documentclass[a4paper]{article}
\usepackage[usenames,dvipsnames]{xcolor}
\usepackage{booktabs}
\usepackage{authblk}
\usepackage{svg}
\usepackage{listings}

\usepackage[hyphens]{url}
\usepackage[colorlinks]{hyperref}
\hypersetup{
	breaklinks=true,
	pdfusetitle=true,
	colorlinks = true,
	linkcolor = blue,
	anchorcolor = blue,
	citecolor = blue,
	filecolor = blue,
	urlcolor = blue
}
\usepackage{breakurl}
\lstdefinelanguage{Ini}
{
	basicstyle=\ttfamily\small,
	columns=fullflexible,
	morecomment=[s][\color{Orchid}\bfseries]{[}{]},
	morecomment=[l]{\#},
	morecomment=[l]{;},
	commentstyle=\color{gray}\ttfamily,
	morekeywords={},
	otherkeywords={=,:},
	keywordstyle={\color{Red}\bfseries},
	numbers=left,
}
\usepackage[T1]{fontenc} \usepackage{ae} \usepackage{aecompl}

\begin{document}

\title{\vspace{-3cm}Controlling and scripting laboratory hardware with open-source, intuitive interfaces: OpenFlexure Voice Control and OpenFlexure Blockly}

\author[1]{Samuel McDermott\thanks{sjm263@cam.ac.uk}}
\author[2]{Richard Bowman}
\author[2]{Kerrianne Harrington}
\author[2]{William Wadsworth}
\author[1]{Pietro Cicuta}

\date{January 6, 2023}

\affil[1]{Department of Physics, Cambridge University, UK}
\affil[2]{Department of Physics, Bath University, UK}
\vspace{-2cm}
\maketitle

\begin{abstract}
Making user interaction with laboratory equipment more convenient and intuitive should promote experimental work and help researchers to complete their tasks efficiently. The most common form of interaction in current instrumentation is either direct tactile, with buttons and knobs, or interfaced through a computer, using a mouse and keyboard. Scripting is another function typical of smart and automated laboratory equipment, yet users are currently required to learn bespoke programming languages and libraries for individual pieces of equipment. In this paper, we present two open-source, novel and intuitive ways of interacting with and scripting laboratory equipment. We choose the OpenFlexure family of microscopes as our exemplar, due to their open-source nature and smart control system. Firstly, we demonstrate `OpenFlexure Voice Control' to enable users to control the microscope hands-free. Secondly, we present `OpenFlexure Blockly' which uses the Blockly Visual Programming Language to enable users to easily create scripts for the microscope, using a drag and drop Web interface. We explain the design choices when developing these tools, and discuss more typical use cases and more general applications.
\end{abstract}

\section{Introduction}
All laboratory equipment requires user input. This generally includes setting some measurement parameters, sending commands, and receiving outputs. Some pieces of equipment, for example optical microscopes, have traditionally relied on continuous hands-on interaction from the user. The basic interactive operation of many computerised instruments often uses complex software interfaces, requiring the operator to switch from the lab bench to operating a keyboard. Other equipment, typically spectrometers or plate readers, usually have automated routines corresponding to typical measurements that will perform tasks for minutes or hours. Even the latter, however, are still limited in terms of ease of automation: (a) even if they are scripted using common programming languages, the formats of scripts and libraries are bespoke to each manufacturer or instrument, and (b) for low-end equipment, it is usually assumed that a user will be physically present to position a sample, fix the setting and the measurement, and then recover the data. Needless to say, automated and modular equipment holds the promise of more easily constructing experimental pipelines to deliver higher throughput. Standardisation of such pipelines would lead to more consistent data across laboratories and would most likely open up areas of investigation that might appear unfeasible with manual approaches. However, creating scripts for more complex routines still largely requires users to have extensive software development skills, as well as learning new libraries.  Progress in this direction is not hindered by particular technological obstacles; on the contrary, software and electronics developed in the last decade for other purposes are ripe to be implemented and adapted in the science laboratory.

In the context of custom optical microscopes we have shown that an open-source approach to hardware and control software design can be very effective to rapidly iterate improvements, and to coalesce a community of users and developers who can drive the equipment towards their specific needs. In this paper, we describe two innovative forms of interacting with the family of OpenFlexure Microscopes. Specifically, by demonstrating voice control and an easy to code automation, we aim to showcase how to control microscopes in ways that are user-friendly, intuitive and can allow the scientist to focus attention on alternative aspects of their experiment in the lab. In addition, these technologies provide better accessibility for all researchers. By documenting in detail these approaches, our hope is that it will be possible for interested users to port one or both methods to the control of other equipment as well, and that the community can develop and build open automation standards for a broad range of science lab instrumentation. 

\section{OpenFlexure Microscope}
\begin{figure}
	\centering
	\includegraphics[width=0.8\linewidth]{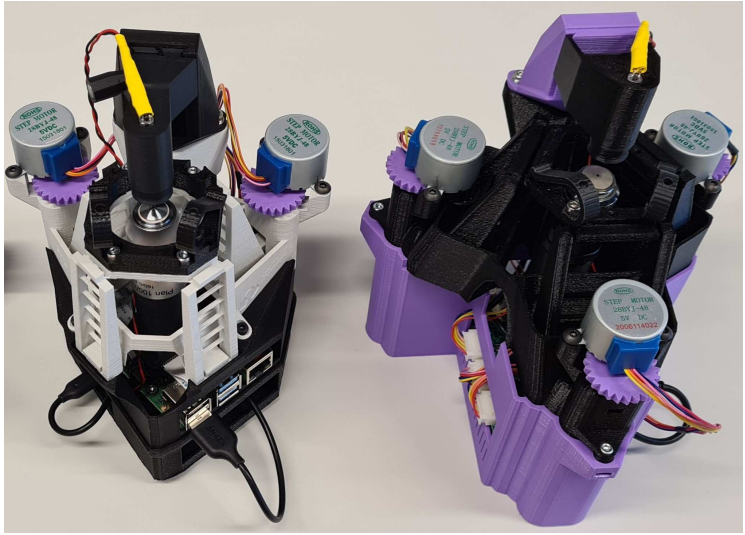}
	\caption{The OpenFlexure Microscope (left) and OpenFlexure Delta Stage (right) are open-source microscopes which are seeing considerable uptake in laboratories around the world~\cite{OpenFlexureForumLocationSurvey}. These devices are used as test devices for the novel laboratory control interfaces presented in this paper.}
	\label{fig:ofm_ds}
\end{figure}

The OpenFlexure Microscope (OFM)~\cite{Sharkey2016,Collins2019RoboticMicroscope} (and similarly its recent development the OpenFlexure Delta Stage (OFDS)~\cite{McDermott2022}), is a low-cost, mostly 3D printed microscope (Figure \ref{fig:ofm_ds}). It is fully open-source, meaning that the parts and assembly instructions can be downloaded.  Any users can then manufacture and modify the microscopes. It is beginning to see considerable uptake, both in research laboratories and educational settings. An informal survey run on the OpenFlexure Forum~\cite{OpenFlexureForumLocationSurvey} has shown that OpenFlexure devices have been built or used in at least 45 countries, in every continent. It is small and affordable enough to be a `personal' and/or `dedicated' microscope, permanently stationed on a lab desk or in a remote location, or in a Microbiological Safety Cabinet (MSC) where one would not want to consign a traditional unit. 

\subsection{Microscope Hardware}
The OpenFlexure family of microscopes has imaging optics and a 3-axis positioning system. They use a Raspberry Pi and a Raspberry Pi camera to capture digital images and an Arduino-based motor control board~\cite{2021SangaboardRepository}. They are discussed in detail in~\cite{McDermott2022}. 

\subsection{Server software}
The OpenFlexure Microscopes are controlled using a web server (written in Python using the Flask framework) running on the embedded Raspberry Pi~\cite{Collins2021}. The API is OpenAPI compliant~\cite{Foundation2022} and follows the W3C web-of-things (WOT) standard~\cite{Kovatsch2020}. 

`Extensions' are software packages that add additional functionality to the microscope, without changing the core code-base. They can be downloaded and installed independently and can provide additional APIs and define Graphical User Interfaces (GUIs). 

\subsection{Clients}
By exposing an HTTP API, clients can send commands to the microscope in a language-agnostic way. Most modern programming languages (e.g. Python) or environments (e.g. MATLAB) have the ability to send HTTP requests.

The most common way that users interact with the OFM is with the built-in web application, providing a Vue-based GUI~\cite{Collins2021}. This web application can be opened as an Electron App or in a web browser. It presents a simple interface to the user with the core functionality of the microscope, as well as extensions. It also displays a live stream from the microscope. OpenFlexure Connect can be used on the Raspberry Pi itself (locally) or on a separate computer (remotely), which is connected to the same network as the microscope. This architecture already allows remote control of the unit in real time.

OpenFlexure Connect works well for sending individual commands, but for more complicated use cases (e.g. long measurements that cycle through parameter sets or even experiments that rely on feedback from image processing or other sensors), scripting is necessary. As the OpenFlexure Software stack can receive HTTP API requests, they can be sent from any language that has that capability. For programmatic scripting, we have developed Python~\cite{2020OpenFlexureRepository} and MATLAB~\cite{2021OpenFlexureClient} clients, which enable users to easily create scripts to control the microscope and receive images back.

\section{Voice Control}
We have designed a voice control client for users who cannot physically interact with the OpenFlexure Microscope (and in future could be adapted to other instrumentation)~\cite{OpenFlexureVoiceControlRepository}. This is particularly suited for users who need to work hands-on with other parts of an experiment, or who have accessibility needs, or for activity in laboratories with substances that can cause harm, so the microscope would be in some enclosure, for example a Microbiological Safety Cabinet (MSC). 

Voice control of electronic devices has become commonplace in domestic and automotive settings, where hands-free control is convenient and important. Well-known brands/environments are Amazon Alexa, Google Assistant, and Apple's Siri voice assistants. Voice control is also starting to make its way into industrial settings~\cite{VoiceCommandIndustry}. However, this important breakthrough in user interaction has yet to gain traction in research laboratories. Most laboratory equipment requires input to be done by hand, by pressing buttons, rotating knobs, or interacting with a computer through a mouse and keyboard. For many applications these inputs are satisfactory, and there can be advantages for example in providing tactile feedback to a user. However, in many circumstances of laboratory work, it would actually be advantageous to reduce the number of objects that are touched. A researcher could be wearing gloves that prevent fine motor action, such as in an MSC, or could be handling hazardous materials, such as chemicals or biological substances, such that all surfaces that are touched need to be cleaned to prevent contamination. In these situations, when equipment is used by various users, the risk of cross-contamination from small mistakes is significant. If users are able to control devices with their voice they are then able to interact with it in a hands-free way which might be important in complex experiments where multiple actions are to be performed at once. 

\begin{figure}[t!]
	\centering
	\includegraphics[width=\linewidth]{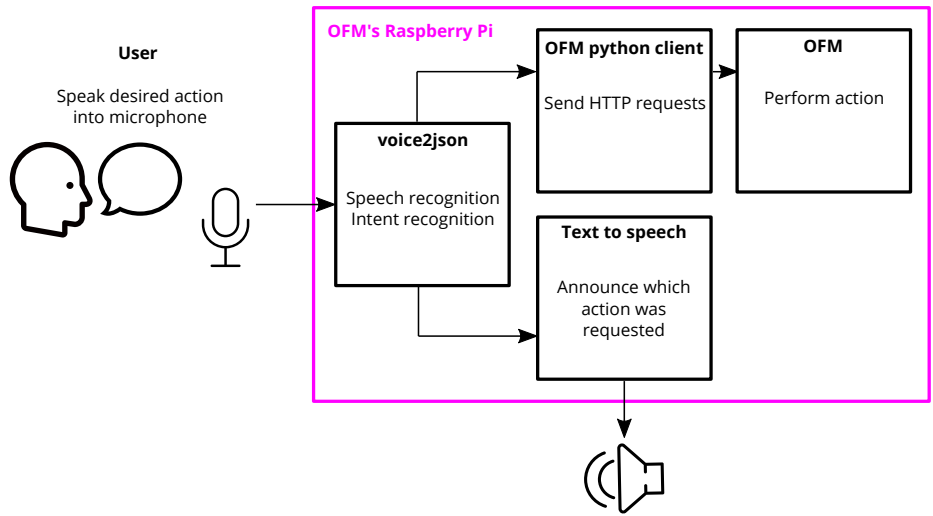}
	\caption{A schematic overview of how we have implemented Voice Control for the OpenFlexure Microscope, using the voice2json library for local processing of voice.}
	\label{fig:voice_schematic}
\end{figure}

In contrast to existing approaches~\cite{Austerjost2018}, we wanted to develop a system that enables voice control of the equipment for the largest possible number of users, and avoid making the laboratory equipment dependent on external proprietary systems. We therefore used an approach which is completely open-source, enabling full transparency. In addition, we aimed to develop a system that would not depend on the cloud, since many potential users of the OpenFlexure Microscope (and other equipment in remote locations) will not have a reliable internet connection at point of use. In addition, we wanted a solution which fulfils important security and privacy concerns, e.g. for researchers working in information-sensitive research.  Both of these conditiions requires that the audio processing happens locally. To meet these requirements, OpenFlexure Voice Control was built using voice2json~\cite{Hansen}, see Figure~\ref{fig:voice_schematic}.

\subsection{Software design}
\subsubsection{The voice2json library}
voice2json is an open-source set of command line tools for offline speech and intent recognition~\cite{Hansen}. It converts audio data in the form of voice commands into JSON events. The process consists of four stages. Speech recognition and intent recognition is handled by voice2json. It is then necessary to convert those intents into commands to send to the OFM. Text-to-speech is then used to communicate back to the user that their command was received.

\subsubsection{Audio Hardware setup}
The only additional hardware required for voice control is a microphone and speaker. Our testing showed that the easiest setup is to use a USB microphone, the majority of which should be plug and play. For the speaker, it is possible to use a bluetooth speaker, the headphone jack or the sound output over an HDMI connection. For busier rooms, we would recommend a bluetooth headset.

\subsubsection{Wake-word}
To prevent unintended commands from being issued, a wake word is used to indicate when the user wants to interact with the microscope. voice2json uses open-source Mycroft Precise~\cite{MycroftAI} as the wake-word listener. This listens to the constant stream of audio from the microphone until it recognises a specific phrase. We used the default phrase ``Hey, Mycroft" because it has already been well trained and tested.  Once it recognises that the phrase ``Hey, Mycroft" has been spoken, a connection to the microscope server is made, and then speech recognition begins.

\subsubsection{Speech recognition}
To be able to interpret what the user wants to do, firstly their speech must be transcribed. voice2json automatically detects when the speech has been started using the webrtcvad library~\cite{Wiseman}. Once detected, it is transcribed using the open source, offline, Kaldi transcriber~\cite{povey2011kaldi}. This transcriber consists of an acoustic model, which maps acoustic and sound features to their corresponding phonemes (small units of speech) and a pronunciation dictionary, which maps combinations of these phonemes to possible words. Finally, the language model orders these possible words by the probability that they will follow each other. This is done using heuristics. 

voice2json uses a custom `sentences.ini` file to generate a custom language model using \textit{opengrm}~\cite{2022OpenGrmLibrary}. This means that the language model can be trained in a short amount of time, approximately 10 seconds on a Raspberry Pi 4 4GB. We created a list of possible sentences that the user could use to control the microscope and stored them in this `sentences.ini` file. In order to reduce the number of commands that would need to be written in this file, voice2json uses a grammar to produce all possible combinations, such as including optional words or ranges of numbers. These possible combinations can be converted to a finite-state network graph for use by opengrm. 

We require some uncommon words such as `autofocus', for which there is no heuristic entry in the pronunciation dictionary. These words are guessed with a `Grapheme to Phoneme model', \textit{Phonetisaurus}~\cite{2022PhonetisaurusRepository}. For autofocus, the predicted phonemes are \texttt{'O t o U f 'o U k V s}. 

Once the sentence has been transcribed and if it matches one of the possible custom sentences, then it is output in JSON for intent recognition. If it does not match one of the possible sentences, then the speech recognition unit continues listening for the next command.

\subsubsection{Intent recognition}
Each one of the possible custom sentences listed in the `sentences.ini` file is stored under a heading, or \textit{intention}. Once the transcriber has determined which sentence was spoken, the intent recogniser looks at the same list of sentences to determine what the intention of that statement would be. It is also possible to add `meta' words into the sentences, which are turned into variables. For example, a common task is to move the stage. This has the intention `MoveStage`, with the following sentence grammar:

\begin{lstlisting}
	[MoveStage]
	axes = (x | y | z){axis}
	relative = (by:True | to:False){relative!bool}
	steps = (-10000..10000){steps}
	move <axes> [axis] <relative> <steps> [steps]
\end{lstlisting}

In line 1, \lstinline![MoveStage]! is the intention name, which will be output in JSON. Lines 2-4 declare the `meta' words, which will be passed as parameters with the names in curly brackets. Line 2 describes the axes, which can be ``\lstinline!x!'' or ``\lstinline!y!'' or ``\lstinline!z!''. Line 3 describes whether the movement is absolute or relative. If the sentence contains ``\lstinline!by!'', then the movement is relative. This parameter is also converted to a Boolean parameter. Line 4 describes a numerical range for the amount of steps to move, in this case between -10000 and 10000. Finally, line 5 constructs the sentence. The words in angle brackets can take any of the values in the `meta' words. The words in square brackets are optional. For example, if a user said the following sentence: `move x axis by 100', it would match this sentence and therefore the ``\lstinline!MoveStage!'' intention. The parameters ``\lstinline!axis = x!'', ``\lstinline!relative = True!'', and ``\lstinline!steps = 100!'' would be output in the JSON alongside this intention as `slots'. It is therefore clear to see how more complicated sentences could be constructed to closely match what we expect the end-users to say. 

\subsubsection{OpenFlexure commands}
Having received a JSON output from the intent recogniser, we can then process the command. We match the intention name directly to a function and use the intention slots as parameters. To send a command to the OpenFlexure Microscope server, we use the previously developed OpenFlexure Python Client~\cite{2020OpenFlexureRepository}. For a given intention, we use the OpenFlexure Python Client to send an HTTP command to the internal OpenFlexure server, which then executes the command.

\subsubsection{Text to speech}
To provide a confirmation to the user that the command was correctly interpreted, eSpeak NG~\cite{2022ESpeakRepository} is used to repeat the recognised voice command back to the user.

\subsubsection{End listening}
Once the OpenFlexure Voice Control is in listening mode, it is possible to send multiple voice commands, one after another. To ensure the device does not accidentally receive commands when not needed, we added a command to `stop listening'. Once it has received this command, OpenFlexure Voice Control switches back to listening for the wake word.

\subsection{Installation and training}
Installation is done on the Raspberry Pi in the OFM. Once the repository has been downloaded, a bash script can be executed, which performs the following steps:

\begin{enumerate}
	\item Get and install voice2json from their repository.
	\item Install \textit{eSpeak NG}, an open source speech synthesizer.
	\item Download the voice2json language profile (english).
	\item Move the `sentences.ini` file (which contains the set of custom template sentences) to the correct location.
	\item Train the voice2json profile. This is done offline, on the Raspberry Pi.
	\item Install the OpenFlexure Python Client.
\end{enumerate}

\subsection{Typical examples of user-equipment interaction}
\begin{figure}
	\centering
	\includegraphics[width=\linewidth]{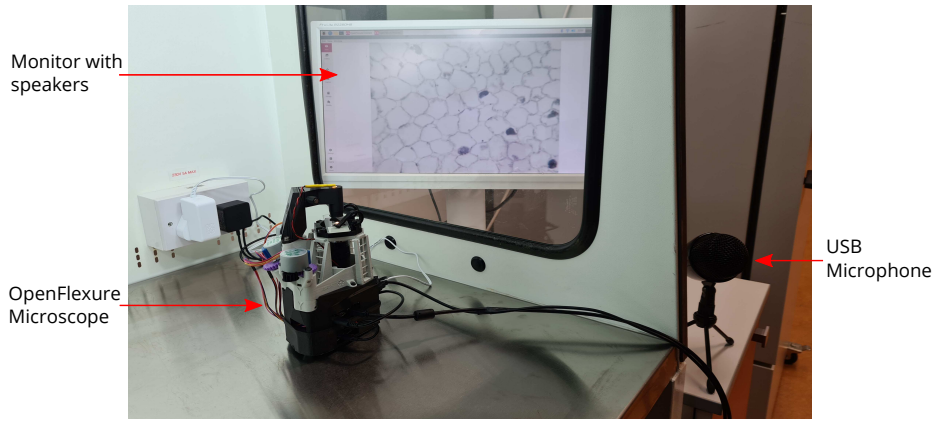}
	\caption{An example of using the OpenFlexure Voice Control in a laboratory environment. The OpenFlexure Microscope, stored here within the laminar flow hood, is connected to a microphone by USB and to a monitor with an HDMI cable. The monitor is capable of displaying the screen and play sounds through its speaker. A user here could therefore keep their sample and hands inside the hood whilst controlling the microscope using their voice.}
	\label{fig:voice_equipment}
\end{figure}

We envisage that this system could be used in a laboratory environment where direct contact with input devices such as keyboards or mice should be limited. In 
Figure~\ref{fig:voice_equipment} we demonstrate this setup. The OFM remains within a laminar flow hood (sterile, biological containment). As a user prepares samples, they can place them directly onto the microscope for imaging. The Microscope has a USB microphone which is placed outside the hood. It also has a monitor outside the hood, connected with an HDMI cable. The monitor can output sounds from the microscope and displays the OpenFlexure Connect GUI. The user can see what the microscope is imaging through the side window. The user can then keep their hands within the cabinet and speak commands to the microscope. To start the microscope's active listening, the user says `Hey, Mycroft'. Then `Autofocus', which triggers the microscope's automated autofocus regime~\cite{Knapper2022FastMicroscope}, and the sample will now be in focus. The user could then say `Capture image', and a digital image of the sample would be stored for later analysis. The user might then want to look at a different region, and could achieve this by saying `Move x-axis by 1000 steps'. Once they have imaged their sample sufficiently, they can say `Stop listening', and the microscope will stop active listening. 

\subsection{Discussion}
Here we have demonstrated how it is possible to use a completely open-source chain of libraries and programs to provide useful voice interaction to the OpenFlexure Microscope without accessing the cloud. This model of voice control could be applied to other devices in a laboratory ecosystem.  

\subsubsection{Automated API}
One of the largest overheads in developing a voice control system for a piece of laboratory equipment is to define the sentences and commands. We have taken care to choose the core set of commands which a user is likely to need when operating the microscope hands-free. A future development could use a standard API specification to automate the process of generating voice commands. For example, the OpenFlexure software is compatible with the Thing Description, from the W3C standard~\cite{W3C2020}, which generates machine-readable interface files for describing the RESTful web service. However, care would need to be taken in curating the generated commands, firstly to make sure that users cannot inadvertently make developer-level changes to their device which they did not intend. Second, there is a risk that too many commands will be created, most of which are not required. This can introduce uncertainty in the voice control logic as to which command the user intended. 

However, one area where automated APIs would be beneficial is for users who are developing their own extensions outside of the core set of commands. As the extensions tend to have fewer commands, automated generation of voice commands from their Thing Description would be a convenient way of extending the range of voice commands. 

\subsubsection{Privacy considerations}
Naturally, some users may have privacy concerns with using voice recognition. voice2json is an edge speech recognition program. Once downloaded, it no longer requires an internet connection. No voice data is ever stored, nor any abstraction of that data. Once the model is trained using the sentences text file, no more training is done. User activity is not recorded. Because all the components of OpenFlexure Voice Control are completely open source, it is possible to independently verify this.

\subsubsection{Internationalisation}
OpenFlexure microscopes are used around the world. As such, for many users, English may not be their preferred language. It is easy to add internationalisation to OpenFlexure Voice Control, and is something we will curate over time with the support of users. To do this, users would need to download the relevant language profile from voice2json. voice2json currently supports the following 18 languages and locales: Catalan, Czech, Dutch, English, French, German, Greek,  Hindi, Italian, Kazakh, Korean, Mandarin, Polish, Portuguese, Russian, Spanish, Swedish, and Vietnamese. Developers would only need to translate the `sentences.ini' for each language, but keep the intent headings the same. The different voice2json language profiles will still manage the transcription and intent recognition.   

\section{Graphical scripting}
Scripting is an essential feature of a smart microscope. The ability to control the device in an ordered way is useful for functionality such as time lapses, region-of-interest imaging, tiling or z-scans, and is a prerequisite for more complex feedback-controlled automated functionality. For this, we have developed Python~\cite{2020OpenFlexureRepository} and MATLAB~\cite{2021OpenFlexureClient} clients. However, many users of the microscope need to link different actions together without writing code, because they might not be familiar with text-based programming or don't have time to learn a new library. To solve this, we present here an implementation with the Visual Programming Language (VPL) ``Blockly''~\cite{2022Blockly, Fraser2015, Pasternak2017}. This open-source visual programming language, developed by Google, enables users to generate syntactically correct code using drag and drop `blocks'. In addition, it is a tool that is widely used in educational settings to teach the fundamentals of coding. Blockly is used by hundreds of projects, including the popular Scratch educational tool. We envisage that the Blockly VPL will enhance the OpenFlexure Microscope as an interdisciplinary education tool. We developed a web app and an extension using the Blockly VPL, with custom blocks for the core controls of the OFM~\cite{OpenFlexureBlocklyRepository}.

\subsection{Functionality}
Google's Blockly is a JavaScript based tool. Developers can design blocks using the developer tools~\cite{2022BlocklyTools} and write the code to which each block corresponds. For example, a block that has the label `capture image' will have code that sends an HTTP request to the OFM to capture an image. Blockly also allows developers to define which inputs can be used for a block (for example, a number input or a drop-down). They can also specify the connections that each block can make to other blocks (for example, if it is an output block). Tooltips can be defined for each custom block, providing hints about its functionality. Based on research by the Blockly development team, they have condensed the lessons they have learned about developing blocks for new users in their style guide~\cite{Fraser2015}. They have already created a collection of pre-written code blocks which correspond to key programming concepts such as `for-loops' and `if-statements'. The developer then compiles a toolbox of the blocks that they wish to include in their app. Blockly itself does not execute any code, instead it generates a list of syntactically correct code, in the language of your choice. For our use case, we chose JavaScript because it is possible to run in the browser. 

Making API calls using Javascript and Ajax is asynchronous. A request is made to the server (in this case, the OpenFlexure microscope), the server processes the result, and then returns a result. Once the result is returned, it is passed to a callback function. For a scripting scenario, this is not suitable, since all the requests will be sent simultaneously rather than one at a time. We used \textit{JS-Interpreter}~\cite{JS-InterpreterRepository}, a JavaScript library designed for use with Blockly, which provides a sand-boxed JavaScript interpreter that allows line-by-line execution of JavaScript code and wraps asynchronous code to enable the script generated by Blockly to run synchronously.

\begin{figure}[t!]
	\centering
	\includegraphics[width = \linewidth]{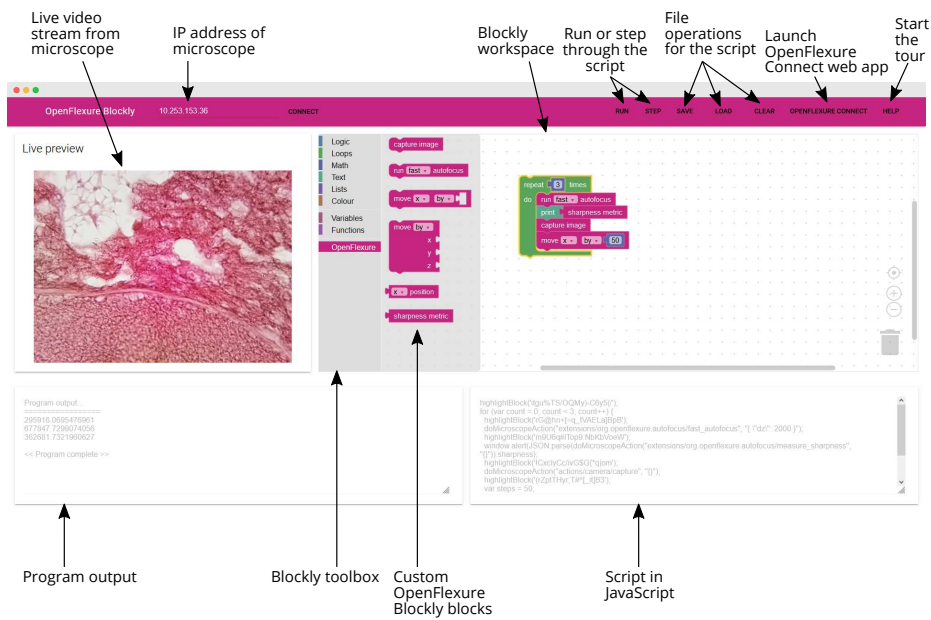}
	\caption{The layout of the OpenFlexure Blockly web app was designed such that all the information is present on one screen. In the top left, users can connect to their microscope with its IP address. The menu bar consists of the operations that the user can do, such as running the script, or stepping through the script line-by-line. It also contains the file operations and links to launch the OpenFlexure Connect web app and start the `help' tour. The live preview box contains the live video stream from the microscope. The Blockly box contains the Blockly workspace and toolbox. At the bottom of the screen is the program output box, containing the statements printed by the script, and the JavaScript box, which displays the JavaScript commands generated from the Blockly script. }
	\label{fig:webapp_design}
\end{figure}

\subsection{Web app}

The web app provides a intuitive interface for scripting the microscope. Users are presented with a simple one-page interface (Figure \ref{fig:webapp_design}) with four boxes. On the top of the window is a menu bar with the buttons for interacting with the app. They are prompted to connect to the microscope, via the microscope IP address, or using an mDNS hostname, by default \texttt{microscope.local}. If the web app is being used on the microscope itself, then it will connect to the microscope automatically. Once connected, a live feed from the microscope's camera is shown in the first box. In the next box, the Blockly graphical interface is shown. This consists of the standard Blockly blocks, as well as the OpenFlexure ones. Users can then drag and drop the blocks from the toolbox into the workspace to create their scripts. The bottom two boxes display the output from the program and the JavaScript that is produced by the blocks.

The web app can be downloaded to be run on a local computer, or can run on the OFM's Raspberry Pi. Once downloaded, the computer or OFM does not need to be connected to the internet, but they do need to be on the same network. Additionally, we host the web app using GitLab's `Pages' functionality. To use this, no installation is required, but it is recommended to use the Firefox browser.

\subsubsection{Menu bar}
At the top of the web app is a menu bar. This contains all the buttons needed to interact with the app. Users can input the IP address of the microscope they are connecting to and connect to it. There are two buttons for running the users script. The \texttt{RUN} button creates the JavaScript code from the blocks that the user has arranged in the workspace. It then uses JS-interpreter to run through the commands continuously, one after another. The \texttt{STEP} button again generates the JavaScript code, but this time the JS-interpreter runs through the code and stops after every command. In this mode, users can see what is happening when they run their code. The \texttt{SAVE} and \texttt{LOAD} buttons allow users to save and load their scripts in a file, so that they can be reused. The scripts are saved in Blockly XML format. The \texttt{CLEAR} removes all blocks from the workspace for a fresh start. The \texttt{OPENFLEXURE CONNECT} button opens the standard OpenFlexure web GUI in a new window, so that users can control the microscope, for example, to get it to the correct starting position. The \texttt{HELP} button launches a step-by-step on-boarding tour, powered by \textit{Intro.js} ~\cite{2022Intro.js}. 

\subsubsection{Microscope live feed}
The OFM generates an MJPEG video stream whenever the server is on. This is used by the clients to display the stream directly from the camera in their GUI. It is inserted into the web app so that users can see their code in action during execution.

\begin{figure}[t!]
	\centering
	\includegraphics[width=\linewidth]{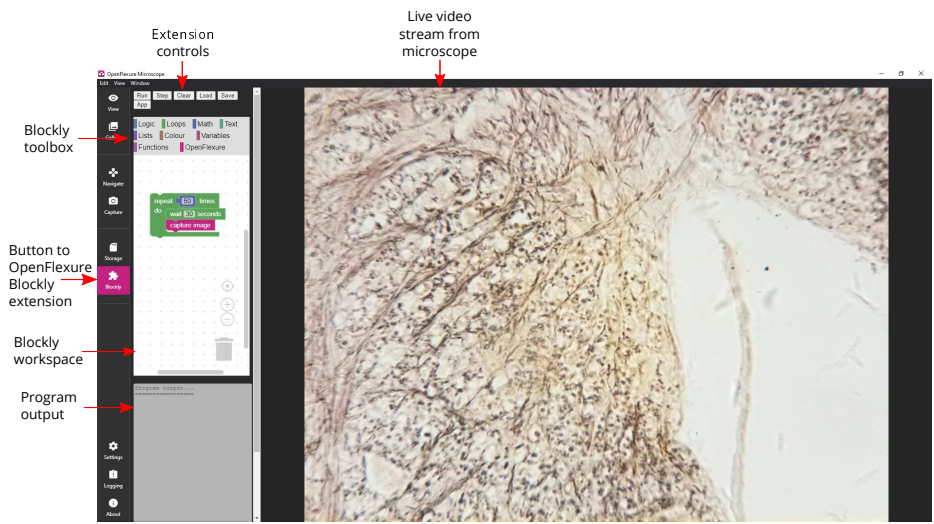}
	\caption{OpenFlexure Blockly can also run as an extension within the OpenFlexure web app, OpenFlexure Connect. This integration allows users to write and run scripts from within the standard web interface GUI.}
	\label{fig:extension_design}
\end{figure}

\subsubsection{Blockly workspace}
The standard Blockly workspace graphical interface is used due to its robustness and familiarity with users. Using the Blockly JavaScript library, the workspace is injected into the web page. It is also given a list of options, such as which renderer to use and which User Interface (UI) components to include. Furthermore, the \texttt{toolbox} is included, which lists the blocks (for our application, all standard blocks are included, as well as our application-specific OpenFlexure Blocks. Users can drag and drop blocks from the toolbox into the workspace and create their scripts using a combination of the standard Blockly blocks and the OpenFlexure blocks. Users can drag, rearrange, and delete blocks as they wish. Blockly's own logic restricts where blocks can be placed so as to make syntactic sense (for example, it is not possible to place a block with no number output in the number of times a for loop should be repeated).  

\subsubsection{Program output}
Users are able to use \texttt{print} statements in their script. They could use this to output the current position of the microscope stage, or the focus metric. Any outputs that they print will be written in this box.

\subsubsection{JavaScript code}
When users click on \texttt{RUN} or \texttt{STEP}, syntactically correct JavaScript is produced from their blocks. The generated JavaScript is displayed in this box. This fulfils two purposes. Users could run this script independently of OpenFlexure Blockly. As they become more confident in programming scripts, they can investigate what the underlying code is, and start to write their own. Generating JavaScript code was chosen because it can be run in the browser. However, Blockly allows for output in several different languages, so in future iterations OpenFlexure Blockly could generate syntactically correct Python code which users could run independently.

\subsection{Running OpenFlexure Blockly as an `Extension'}
As users become more familiar with the OFM, we hope that they will become confident in using OpenFlexure Blockly and OpenFlexure Connect. Therefore, we have also created an extension that allows them to run OpenFlexure Blockly from within OpenFlexure Connect (Figure \ref{fig:extension_design}). Our recommendation is that users create their OpenFlexure Blockly Script in the OpenFlexure Blockly web app. They can then save this script and reopen it in the OpenFlexure Blockly extension. They can then modify and run their script within OpenFlexure Connect, alongside its more advanced functionality.

\begin{figure}[!t]
	\centering
	\makebox[\textwidth][c]{\includegraphics[width=1.0\linewidth]{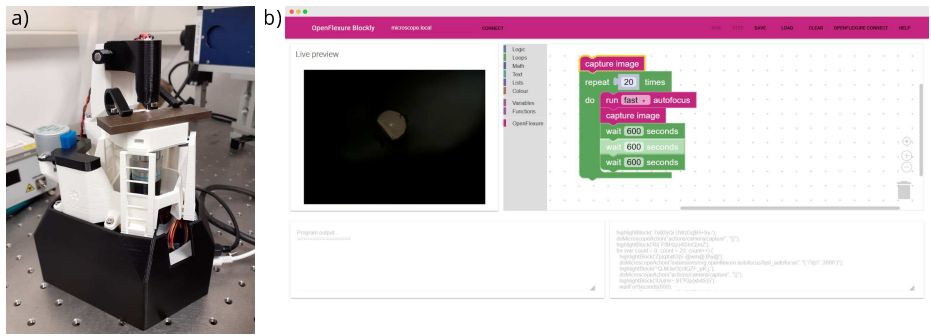}}
	\caption{The OpenFlexure Microscope was used to monitor optical fibre cleaves at regular intervals. (a) A cleaved piece of hollow core fibre sat in a metallic v-groove on the OpenFlexure Microscope stage. (b) The OpenFlexure Blockly web app was used to design a script which autofocuses and take images every 30 minutes over several days.}
	\label{fig:fibre}
\end{figure}

\subsection{Typical user interactions}
We have considered OpenFlexure Blockly for two use cases. The first is for use as an educational tool, primarily for programming beginners. The second is for researchers who quickly want to develop scripts to fulfil more complex operations than are possible with a point and click GUI.

\subsubsection{Education}
The OFM is suitable for education worldwide, and there have been projects using it in both primary and higher education. Its low cost, open-source nature can be attractive to many educational establishments. It requires no user installation, and the web interface can be displayed on a large screen for participants to view. The drag and drop interface will be familiar to many students with the increasing adoption of VPLs for programming education. Students are able to quickly create automated experiments, for example time-lapses, tiling, and repeatable position imaging.

\subsubsection{Research}
OpenFlexure Blockly is also suitable for researchers who wish to quickly make scripts for the OFM. Although we have developed clients in both MATLAB and Python, many researchers will be unfamiliar with these tools, and the barrier to entry could be too high for some researchers who wish to do some simple scripting. OpenFlexure Blockly, which runs on all common web browsers, requires no installation or environment setup. A script can be built quickly, without remembering complicated syntax. The script can easily be saved and loaded to allow for repeatable experiments. 

\subsubsection{Example: Imaging hollow core optical fibre}
OpenFlexure Blockly was used to easily automate the monitoring of optical fibre cleaves over time. This is of particular research interest for hollow core optical fibres as contaminants growing on the end face will reduce the transmission from the input and outputs of bare fibre. It is useful to observe how these contaminates appear on freshly cleaved optical fibre end faces at regular intervals. 

Various hollow core fibres fabricated at the University of Bath were cleaved with a Fujikura CT-101 fibre cleaver, so that `fresh' and low angle end faces were produced. These were loaded onto a metallic v-groove on the OFM stage (Figure \ref{fig:fibre} (a)). The lengths of these cleaved fibres are several millimetres. OpenFlexure Blockly was used to create a script to autofocus and then save the current image, every 30 minutes over several days (Figure \ref{fig:fibre} (b)). Contrary to what has been previously observed~\cite{Rikimi2020GrowthFibers}, no significant growth of similar contaminants was observed on any of the hollow core fibre end faces.

\section{Conclusion}
It is necessary to interact with laboratory equipment in order to provide it with inputs and receive outputs. Typically, this is through physical components such as buttons, keyboards, and mice. In this paper, we discuss two novel ways that we have developed to interact with laboratory equipment, using the OpenFlexure family of microscopes as example devices. The focus of these developments is to improve the accessibility of the microscope, increasing its potential to more users.  

The OpenFlexure Microscope (OFM) and OpenFlexure Delta Stage (OFDS) are high-functionality, mostly 3D-printed microscopes. They have a 3-axis stage which is moved by motors, and they capture digital images using a Raspberry Pi and Raspberry Pi camera. Because they are small, open source, and low cost they are a useful tool for researchers and educators. They can be are equally at home in MSCs, imaging live samples, or in classrooms, educating the next generation of microscopists. Thanks to the server-client architecture of the OpenFlexure software, it is possible to create different clients that can control the microscope.  

Advanced microscopes typically require a mouse, keyboard, joystick, or other kind of physical input control. This presents containment issues, normally requiring users to change gloves whilst controlling this equipment. These input devices are also difficult to clean and are not suitable for users with accessibility issues. Therefore, we developed OpenFlexure Voice Control, building upon open source, edge computing speech and intent recognition libraries. We explained how OpenFlexure Voice Control was built and demonstrated a potential use case in a laboratory environment.

Designing scripts for advanced microscopes typically require a high level of technical skill. For many programming beginners, this is a daunting task, especially when they are more interested in how their sample looks. We therefore developed OpenFlexure Blockly, a tool to create scripts for the OpenFlexure family of microscopes using Google's open source Visual Programming Language (VPL), Blockly. This allows users to create scripts by dragging and dropping blocks, producing syntactically perfect JavaScript. The user can then run the script to control the OFM. Two use cases of this were provided, one for education where the principles of microscopy can be taken to a higher level with more complex imaging routines. It was also shown to be useful for researchers who wanted to quickly create scripting routines to image their samples.

This paper illustrates how by moving beyond the conventions of interaction with smart microscopes we can increase their usability and accessibility. Our work demonstrates how to add these novel UI experiences with open source libraries, and with the increase in smarter equipment, we hope these approaches will be taken up across other laboratory instrumentation. 

\section{Data Availability}
All source code is open-source and available on our repositories:
\begin{itemize}
	\item OpenFlexure Voice Control Repository~\cite{OpenFlexureVoiceControlRepository}
	\item OpenFlexure Blockly Repository~\cite{OpenFlexureBlocklyRepository}
\end{itemize}

\section*{Acknowledgements}
Funding was received from the following grants:   EPSRC: EP/R013969/1, Royal Society: URF\textbackslash R1\textbackslash 180153, Royal Academy of Engineering: ING1920\textbackslash14\textbackslash39.   We thank Dr Joel Collins for the original suggestion of using Blockly.

\bibliographystyle{IEEEtran}
\bibliography{main}
\end{document}